\documentclass[aps,prx,superscriptaddress,
  twocolumn,
  citeautoscript,
  longbibliography,
  showpacs,floatfix]{revtex4-1}
\usepackage{amssymb} \usepackage{amsfonts} \usepackage{amsmath}
\usepackage{graphicx} \usepackage{ulem}
\usepackage{color}

\renewcommand{\d}{\text{d}}

\newcommand{\Ft}{\tilde{F}} \newcommand{\chit}{\tilde{\chi}}

\newcommand{\nm}{\text{nm}} 
 \newcommand{\vdW}{\text{vdW}}
\newcommand{\RPA}{\text{RPA}}

\newcommand{\lbr}{\left(} \newcommand{\rbr}{\right)}
\newcommand{\lbrs}{\left[} \newcommand{\rbrs}{\right]}
\newcommand{\lbrc}{\left\{} \newcommand{\rbrc}{\right\}}

 \newcommand{\vq}{\vec{q}}
 
\newcommand{\qbar}{\bar{q}}

\begin{document}
\title{How many-body effects modify the van der Waals interaction
  between graphene sheets}
\author{John F. Dobson}
\affiliation{Queensland Micro and Nano Technology Centre, Griffith
  University, Nathan, Queensland 4111, Australia}
\affiliation{Donostia International Physics Centre and European
  Theoretical Spectroscopy Centre, Universidad del Pais Vasco, San
  Sebastian, 2008, Spain}
\affiliation{Department of Physics, University of Missouri,
  Columbia MO 65211}
\author{Tim Gould}
\affiliation{Queensland
  Micro and Nano Technology Centre, Griffith University, Nathan,
  Queensland 4111, Australia}
\author{Giovanni Vignale}
\affiliation{Department of Physics, University of Missouri,
  Columbia MO 65211}

\begin{abstract}
  Undoped graphene (Gr) sheets at low temperatures are known, via Random
  Phase Approximation (RPA) calculations, to exhibit unusual van der
  Waals (vdW) forces. Here we show that graphene is the first known system
  where effects beyond the RPA make {\it qualitative} changes to the
  vdW force. For large separations, $D \gtrsim 10$nm
  where only the $\pi_z$ vdW forces remain, we find the
  Gr-Gr vdW interaction is substantially reduced from the RPA
  prediction. Its $D$ dependence is very sensitive to the form of the
  long-wavelength many-body enhancement of the velocity of the massless
  Dirac fermions, and may provide independent confirmation of the
  latter via direct force measurements.
\end{abstract}
\date{\today} \maketitle

\section{Introduction}

It is well known that a zero-gap conical $\pi_{z}$ electronic Bloch
band structure of undoped graphene, supporting massless Dirac fermions
propagating with speed $v$, should give this system a number of unusual
properties.~\cite{Neto2009} One such property relates to the
low-temperature dispersion (van der Waals) interaction energy per unit
area $E^{\vdW}/A$ between undoped parallel graphene sheets separated
by a large distance $D$. Commonly used theories such as the summation
of pairwise atomic contributions, $E^{\vdW}\approx\sum_{i\neq
  j}C_{ij}R_{ij}^{-6}$, and other popular and largely successful
approaches~\cite{Grimme2006,Dion2004,Langreth2005,%
  Tkatchenko2009,Vydrov2010}, predict the energy for this
case (and any case with parallel 2D sheet geometry) to be a power law
\begin{align}
\frac{E^{\rm vdw}}{A}=&-\frac{C_{4}}{D^4}
\label{DMinus4Law}
\end{align}
where $C_{4}$ is a system-dependent constant (see Sections 4 and 8 of
Ref.~\onlinecite{Dobson2012-JPCM} for further discussion). By
contrast, more microscopic/collective approaches, such as the RPA
correlation energy based on the graphene $\pi_{z}$ -$\pi_{z}^{\ast}$
electronic response, yield the
result~\cite{Dobson2006,Gould2008,Sernelius2012,Gould2013-Cones,Sharma2014}
\begin{align}
\frac{E^{\rm vdw}}{A}=&-\frac{C_{3}}{D^3}\,.
\label{DMinus3Law}
\end{align}
The constant $C_3$ is easily calculated within the random phase
approximation (RPA), which treats the electrons in each layer as
essentially non-interacting, but subjected to their own
time-dependent classical
electrostatic field.  Indeed, if one makes the simplest
(``Casimir-Polder'') approximation, in which the interlayer Coulomb
interaction is treated by second-order perturbation theory, one finds,
in the limit of large separation
\begin{align}
C_3 =& \frac{e^2}{32\pi} F\left(\frac{\pi}{2}\alpha\right)\,,
\label{C3Formula}
\end{align}
where
$\alpha = \frac{e^2}{\hbar v}$ is the effective
fine structure constant of
graphene, related to the velocity $v$ of the massless Dirac fermions
as the fundamental fine structure constant is to the speed of light,
and $F(a)$ is a smoothly varying function,
which is plotted in Fig.~\ref{Fig1}
(the derivation of this result, as well as the
analytic expression for $F(a)$ will be presented below).
\begin{figure}
\includegraphics[width=1.0\linewidth]{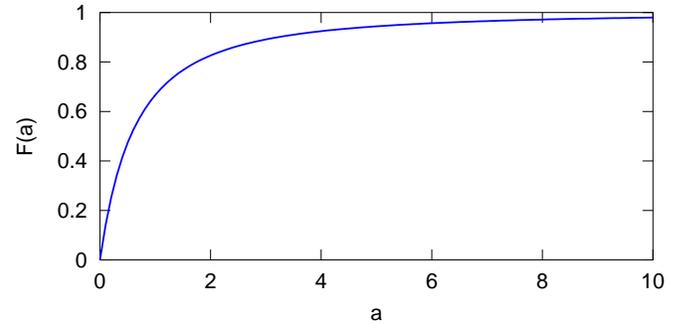}%
\caption{Plot of the function $F$ which controls, via
  Eq.~\eqref{C3Formula}, the value of the constant
  $C_3$ for the RPA van der Waals interaction between graphene sheets.}
\label{Fig1}
\end{figure}
For $v=10^6$m/s one has $\alpha \simeq 2.2$, which makes $C_3$ very
close to the ``universal" value
$C_3^{\rm uni} = \frac{e^2}{32\pi}$. This value is
weakly dependent on small variations of $v$, and remarkably,
the interaction $-C_3^{\rm uni}/D^3$ is not changed by the inclusion of
electromagnetic retardation, even at large $D$ values.
(see Eq.~36 of \cite{Drosdoff2010}, \cite{Drosdoff2012}, and
Eq.~8b of \cite{[][Note: {$\chi_0$} in this work was too large
by a factor of two]Dobson2006} corrected for a spurious factor 2)

In this context it is important to note that for real graphene
Eq.~\eqref{DMinus3Law} is valid only at large separations: at shorter
distances, gapped transitions other than the
$\pi_{z}\rightarrow\pi_{z}^{\ast}$
contribute a vdW energy of the conventional form \eqref{DMinus4Law}.
It is only for $D\gtrsim 10\nm$ that numerical work within the
RPA\cite{Lebegue2010} suggests the $D^{-3}$ falloff overtakes the
conventional $D^{-4}$ contribution.
It is therefore in this regime of larger separations
corresponding to small wavevectors that one should
check for any many-body effects beyond the RPA due to the
$\pi_{z}-\pi_{z}^{\ast}$ graphene response.
Furthermore at these separations all electrostatic and
metallic overlap forces have long vanished.

\section{Renormalization of the velocity}

Such corrections are worth investigating because the $\pi_{z}$
electrons in a graphene sheet are obviously not ``weakly
interacting'': the coupling constant $\alpha$ is larger than
one\cite{Gonzalez1994,Gonzalez1999,Li2008,Elias2011,Siegel2011}.
What is believed to be true is that the {\it effective} long-wavelength
Hamiltonian, generated by a renormalization group (RG) flow, is
non-interacting.\cite{Gonzalez1994,Gonzalez1999,%
Herbut2006,Mishchenko2009,Kotov2012} However, the reason for this
simplification is not that the electric charge renormalizes to zero
but that the fermion velocity grows to infinity.  More precisely, if
one introduces an effective Fermion velocity $v_q$, which describes
the system at length scales larger than $1/q$, then $v_q \to \infty$
for $q \to 0$, and, accordingly, the running coupling constant
$\alpha_q = \frac{e^2}{\hbar v_q}$ tends to zero.
This long-wavelength  renormalization is quite distinct from
beyond-RPA corrections at short wavelength, arising from modified
forms of  Adiabatic Local Density Functional Theory (ALDA), which are
also sometimes also described as "renormalized"\cite{Olsen2012}
Our long-wavelength corrections can
change the asymptotic power law for the van der Waals interaction,
whereas the short-ranged ALDA-based ones, unsurprisingly, have no
major qualitative effect on long-wavelength vdW
phenomena\cite{Olsen2013-rALDA}.

The stronger the
original interaction is at the microscopic scale, the larger the
renormalized velocity becomes at any given length scale $1/q$.  The
exact form of the divergence of $v_q$ for $q \to 0$ is, of course,
unknown, since the many-body problem has not been solved.
RG calculations based on first-order perturbation
theory~\cite{Gonzalez1994,Kotov2012} suggest a weak logarithmic
divergence of the form
\begin{align}
  v^{(1)}_{q} =& v \lbr
  1+\frac{1}{4}\alpha \ln\frac{\Lambda}{q} \rbr\,,
  &
  \alpha_q^{(1)} =& \frac{\alpha}{1+\frac14\alpha\ln\frac{\Lambda}{q}}\,,
  \label{V1}
\end{align}
where $\Lambda$
is a microscopic cutoff of the order of the inverse of the lattice
constant (\AA).  More sophisticated calculations at the ``two-loop"
level and in the large-$N$
limit,~\cite{Gonzalez1999,Polini2007,DasSarma2007,Son2007,%
  Foster2008,Kotov2009,Kotov2012}
$N$ being the number of Fermion flavors, predict a stronger power-law
divergence of the form
\begin{align}
  v^{(2)}_{q} =& v\left(\frac{\Lambda}{q}\right)^\beta\,,
  &
  \alpha_q^{(2)} =& \frac{\alpha}{\left(\Lambda/q\right)^{\beta}}\,,
  \label{V2}
\end{align}
where $\beta = \frac{8}{N\pi^2}$, ($N=4$ for graphene). 
Recent experiments performed by a
variety of techniques have at least partially confirmed these
theoretical predictions\cite{Li2008,Reed2010,Elias2011,Siegel2011},
showing that the effective coupling constant is reduced and the Dirac
cones are strongly compressed~\cite{Elias2011} near the Dirac point.

The many-body enhancement of the fermion velocity has also been shown
to affect various many-body phenomena, such as the plasmon
dispersion~\cite{Abedinpour2011,Levitov2013}, the optical Drude
weight,~\cite{Abedinpour2011} and the electronic screening of external
charges.\cite{Sodemann2012}
To date\cite{Olsen2012,Olsen2013-rALDA},
beyond-RPA effects were believed to
alter at most the prefactor, not the power exponent, of vdW decay with
distance.  In this Letter we expose a striking case where beyond-RPA
many-body renormalization  affects the essential character of a vdW
interaction, namely that between graphene sheets.
Our main result is that the long-wavelength enhancement of the fermion
velocity causes the vdW interaction to decrease asymptotically faster
than in Eq.~\eqref{DMinus3Law} but still slower than in the
conventional Eq.~\eqref{DMinus4Law}.  This result can be expressed in
an intuitively appealing way by saying that the bare $\alpha$ in
Eq.~\eqref{C3Formula} must be replaced by the running coupling
constant $\alpha_q$ evaluated at $q=1/D$.  Thus we have
\begin{align}
  \frac{E^{\rm vdw}}{A} \simeq -\frac{e^2}{32\pi
    D^3}F\left(\frac{\pi}{2}\alpha_{1/D}\right)\,,
  \label{DMinus3NewLaw}
\end{align}
and since $F(x) \simeq \frac{\pi}{2} x$ for $x \ll 1$ the
asymptotic behavior of the vdW interaction is reduced, relative to the
RPA, precisely by the factor $\alpha_{1/D}$, which vanishes in the
limit $D \to \infty$.  The renormalized interaction will therefore
decrease as $[D^3 \ln (D\Lambda)]^{-1}$ or as $D^{-(3+\beta)}$,
depending on which of the two scenarios,~\eqref{V1} or \eqref{V2}, is
realized.  Additional many-body effects contained in the so-called
vertex corrections turn out to be irrelevant at sufficiently large
distance, even though, of course, they can quantitatively change the
result at intermediate distances. Throughout the analysis, we assume
that the distances are not so large that $v_{q=1/(2D)}$ becomes
comparable to the speed of light, at which point electromagnetic
retardation effects should be taken into account.
Using this criterion, retardation becomes dominant only for $D$ of order
$10^{220}$m (!) for the logarithmic renormalization case
[Eq.~\eqref{V1}] or $O(10^1\text{m})$ for the power law case
[Eq.~\eqref{V1}]. Thus retardation here is unimportant in practice, as
for the pure RPA theory.

\section{van der Waals Calculations}

We consider the interaction energy between two parallel freely suspended
graphene sheets {\it in vacuo}, separated by a distance $D$ that is
much larger than the two-dimensional lattice constant $a$,
as well as the thickness $T$ of each sheet
(see Appendix~\ref{app:Bulk} for other scenarios).
Our starting point is the Casimir-Polder (CP) formula obtained by
doing straightforward second-order perturbation theory in the
inter-layer electron-electron interaction potential
\begin{align}
  V_{\rm inter}(q) =& \frac{2\pi e^2}{q} e^{-qD}\,,
\end{align}
where ${\bf q}$ is the two-dimensional wave
vector of density fluctuations in each layer.  The result is
\begin{align}
\frac{E^{(2)}}{A}=&\frac{-\hbar}{8\pi^{3}}\int_{0}^{\infty}\d u\int
d^{2}{\bf q} \chi(q ,iu)^{2} \lbr \frac{2\pi e^2}{q}e^{-qD}\,.
\rbr^{2}
\label{ECP}
\end{align}
Here $\chi(q,iu)$ is the electronic density-density
response function of a single
isolated layer, evaluated at wave vector ${\bf q}$ and imaginary
frequency $iu$.  The exponentially decaying factor $\exp(-2qD)$
ensures that small wave vectors $q\approx1/D$ dominate
Eq.~\eqref{ECP} for large separations.

The response function $\chi(q,iu)$ is expressed in terms of the proper
response function $\tilde \chi(q,iu)$ and the intra-layer interaction
potential $V_{\rm intra}(q)=\frac{2\pi e^2}{q}$ according to the
well-known formula~\cite{Vignale2005}
\begin{align}
  \chi(q,iu)
  =&\frac{\tilde \chi(q,iu)}{1 - \frac{2\pi e^2}{q}\tilde \chi(q,iu)}
  \label{ChiFull}
\end{align}
In the RPA one approximates $\tilde \chi(q,iu) \approx
\chi_0(q,iu)$, where
\begin{align}
  \chi_0(q,iu)=-\frac{1}{4\hbar v}\frac{q}{\sqrt{1+x^2}}\,,
\end{align}
is the non-interacting
zero-temperature response function for the conical
$\pi_{z}-$bands,\cite{Gonzalez1994,Wunsch2006,Dobson2006}
$v$ is the bare velocity, and $x\equiv \frac{u}{qv}$.

\begin{figure}
\includegraphics[width=1.0\linewidth]{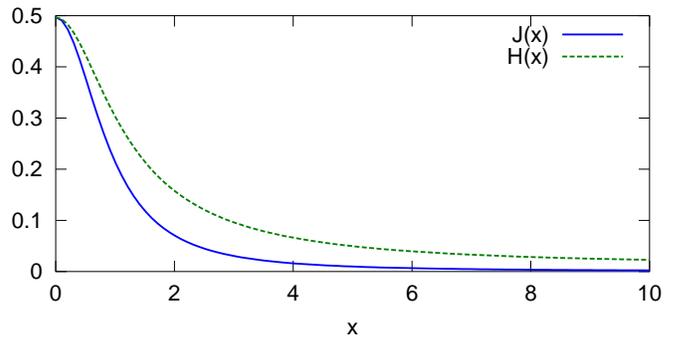}%
\caption{(Color online) Numerical fit [Eq.~\eqref{JFit}] to the
  functions $J(x)$ (solid line) and $H(x) \equiv J(x)\sqrt{1+x^2}$
  (dashed line) which control the magnitude of the self-energy and
  vertex corrections to the RPA beyond the simple renormalization of
  the velocity.}
\label{Fig2}
\end{figure}

Intra-layer electron-electron interactions modify $\tilde \chi$ in two
ways: via self-energy insertions and via vertex corrections.  For
example, in a recent first-order perturbative calculation coupled with
RG arguments, Sodemann and Fogler (SF) find\cite{Sodemann2012}
\begin{align}
  \chit(q,iu)=&-\frac{q}{4\hbar v_{q}} \lbr
  \frac{1}{\sqrt{x^{2}+1}}+\alpha_{q}J( x) \rbr\,,
  \label{FullChTildeScaled}
\end{align}
where $v_q$ is
given by Eq.~\eqref{V1} (i.e., $v_q=v_q^{(1)}$) and $\alpha_q$ is the
corresponding coupling constant.  Here the self-energy insertion has
caused the bare velocity $v$ to be replaced by the renormalized,
scale-dependent velocity $v_q$. The second term, $J$, is a
dimensionless function representing the combined effects of
self-energy and vertex corrections beyond the simple rescaling of $v$.
In the notation of SF $J$ is given by $J( x) \equiv I_{a}( ix) +I_{b}(
ix)$ where the functions $I_{a}$ and $I_{b}$, defined in
Ref.~\onlinecite{Sodemann2012}, are analytically continued here to the
imaginary $x$ axis.  It is essential to our subsequent arguments that
$J(x)$ is a smooth function of $x$, varying monotonically between
$J(0)\approx0.497$ and $J(x\to\infty) \approx\frac{0.013}{x}\equiv
\frac{C_{\infty}}{x}$ (see Eq.~\eqref{JFit} and
Fig.~\ref{Fig2}, where a numerical fit to $J(x)$ is provided).
This feature is expected to persist beyond the first-order
approximation, e.g., even in the strong coupling limit, where the
renormalized velocity is likely given by $v^{(2)}_q$ of Eq.~\eqref{V2}
(see Ref.~\onlinecite{Kotov2012}).  A simple fit to $J(x)$ based on
the results of Ref.~\onlinecite{Sodemann2012} is
\begin{align}
  J(x)=& \frac{(y+a)J(0)+(y^3-1)aC_\infty}{y^3(1+ay)}\,,
  \label{JFit}
\end{align}
where $a = 0.285$ and $y=\sqrt{1+x^2}$.  This is shown in Fig.~\ref{Fig2}. For
imaginary $x$, this numerical fit provides a good match to the results
of Ref.~\onlinecite{Sodemann2012}.

Substituting Eq.~\eqref{FullChTildeScaled} into Eq.~\eqref{ChiFull} we
obtain the full interacting response function in the form
\begin{align}
  \chi(q,iu)=&-\frac{q}{4\hbar
    v}\frac{1+\alpha_qH(x)}{\sqrt{1+x^2} +\frac{\pi}{2}\alpha_q
    [1+\alpha_qH(x)]}\,,
  \label{InteractingChi}
\end{align}
where $H(x)\equiv J(x)\sqrt{1+x^2}$.  Plugging into
Eq.~\eqref{ECP} and changing integration variable from $q$ to $\bar q
= 2qD$ we find, after simple manipulations
\begin{align}
\frac{E^{(2)}}{A}= -\frac{e^2}{32\pi D^3}\int_0^{\infty}
\frac{1}{2}\bar q^2e^{-\bar q} \tilde F\lbr \frac{\pi}{2} \alpha_{\bar
  q/2D}\rbr\d \bar q\,,
\label{ECPIntermediateD}
\end{align}
where we have defined the function
\begin{align}
\tilde F(a) \equiv a\int_{0}^{\infty} \lbrc \frac{1+\frac{2a}{\pi}
  H(x)} {\sqrt{1+x^2} + a[1+\frac{2a}{\pi}H(x) ]} \rbrc^{2}\d x\,.
\label{IntegralForFtilde}
\end{align}
If now self-energy and vertex corrections are neglected by setting
$H=0$, we see that $\tilde F$ simplifies to
\begin{align}
  F(a) \equiv
  a\int_{0}^{\infty} \lbrc \frac{1} {\sqrt{1+x^2} + a} \rbrc^{2}\d x\,,
  \label{IntegralForF}
\end{align}
which can be evaluated analytically, yielding
\begin{align}
F(a)=&\begin{cases} \frac{\frac{a\pi}{2} - a^2\sqrt{1-a^2} -
    a\tan^{-1}\lbr \frac{a}{\sqrt{1-a^2}} \rbr }{(1-a^2)^{3/2}}, & a<
  1 \\ \frac{a^2 -
    \frac{a}{2\sqrt{a^2-1}}\log\frac{a+\sqrt{a^2-1}}{a-\sqrt{a^2-1}}
  }{(a^2-1)}\,, & a>1
\end{cases}
\label{eqn:Ft}
\end{align}
and $F(a)=2/3$ for $a=1$.  This is precisely the function $F$ that was
introduced in Eq.~\eqref{C3Formula}
and was plotted in Fig.~\eqref{Fig1}.
Since in RPA $\alpha_q$ is constant ($\alpha_q=\alpha$) and
$\int_0^\infty \bar q^2e^{-\bar q} \d \bar q
=2$, Eq.~\eqref{C3Formula} is seen to follow immediately from
Eq.~\eqref{ECPIntermediateD}.

To go beyond the RPA we must reinstate the self-energy and the vertex
corrections.  However, we observe that in the large-$D$ limit the
coupling constant tends to zero and therefore the $\tilde F$ function
reduces to the $F$ function, which in turn can be replaced by its
small-$a$ expansion $F(a) \approx \frac{\pi}{2}a$.  Thus,
Eq.~\eqref{ECPIntermediateD} becomes
\begin{align}
  \frac{E^{(2)}}{A} =& -\frac{\pi
    e^2}{256 D^3}\int_0^{\infty}\bar q^2e^{-\bar q} \alpha_{\bar
    q/2D}\d \bar q\,,
  \label{ECPFinal}
\end{align}
The asymptotic behavior of the vdW interaction depends solely on
the behavior of the running coupling constant $\alpha_q$ (or,
equivalently, the renormalized velocity) in the $q \to 0$ limit.
For the logarithmic renormalization case of Eq.~\eqref{V1},
we evaluate \eqref{ECPFinal} by freezing the slowly varying
$\alpha_{q/2D}^{(1)}$, evaluating it at the maximizing value
$q_0=2$ of the rapidly varying integrand, getting
\begin{align}
  \frac{E^{(2)}}{A} \approx&-\frac{\pi e^2}{128 D^{3}}\lbrs
  \frac{\alpha}{1+\frac14\alpha\ln(\Lambda D)}\rbrs\,.
  \label{AsyvdWRedByRenorm}
\end{align}
This shows a modest, logarithmic reduction of the vdW interaction
relative to the RPA result.  If, on the other hand, the
strong-coupling model of Eq.~\eqref{V2} is adopted, \eqref{ECPFinal}
gives an altogether different power-law behavior:
\begin{align}
\frac{E^{(2)}}{A} =-\frac{\pi \Gamma(3+\beta) \Lambda^\beta e^2}{32
  (2D)^{3+\beta}}\,,
\label{AsyvdWRedByRenorm2}
\end{align}
where $\Gamma(x)$ is the gamma function.  Notice that, since $\beta
<1$, this is still larger than the $D^{-4}$ dependence expected for
insulating 2D layers, and therefore dominates at large separations
in real graphene where gapped insulator type transitions also
contribute to the response.

The above eqs.~(\ref{ECPFinal}-\ref{AsyvdWRedByRenorm2}) are valid at
asymptotically large separations.  At finite separations the more
accurate Eqs~\eqref{ECPIntermediateD}-\eqref{IntegralForFtilde} must
be used.  These now depend not only on the fermion velocity -- a
measurable quantity -- but also on the form of the function $J(x)$,
which is not directly accessible to experiment and must be calculated
by many-body theory (Notice, however, that the imaginary part of the
density response function for real frequency is related to the optical
absorption spectrum, which is, in principle, measurable, and could be
used to calculate the function $J$).  Making use of $J(x)$ calculated
in Ref.~\onlinecite{Sodemann2012} and fitted as shown in
Fig.~\ref{Fig2} we find that
\begin{align}
\tilde F(a)\approx \lbr 1 + 0.165\frac{(2.1a)}{\sqrt{1+(2.1a)^2}} \rbr
F(a)
\label{eqn:Fa}
\end{align}
is an excellent approximation (relative error under 1\%) to the
integral of Eq.~\eqref{IntegralForFtilde}.
\begin{figure}
\includegraphics[width=1.0\linewidth]{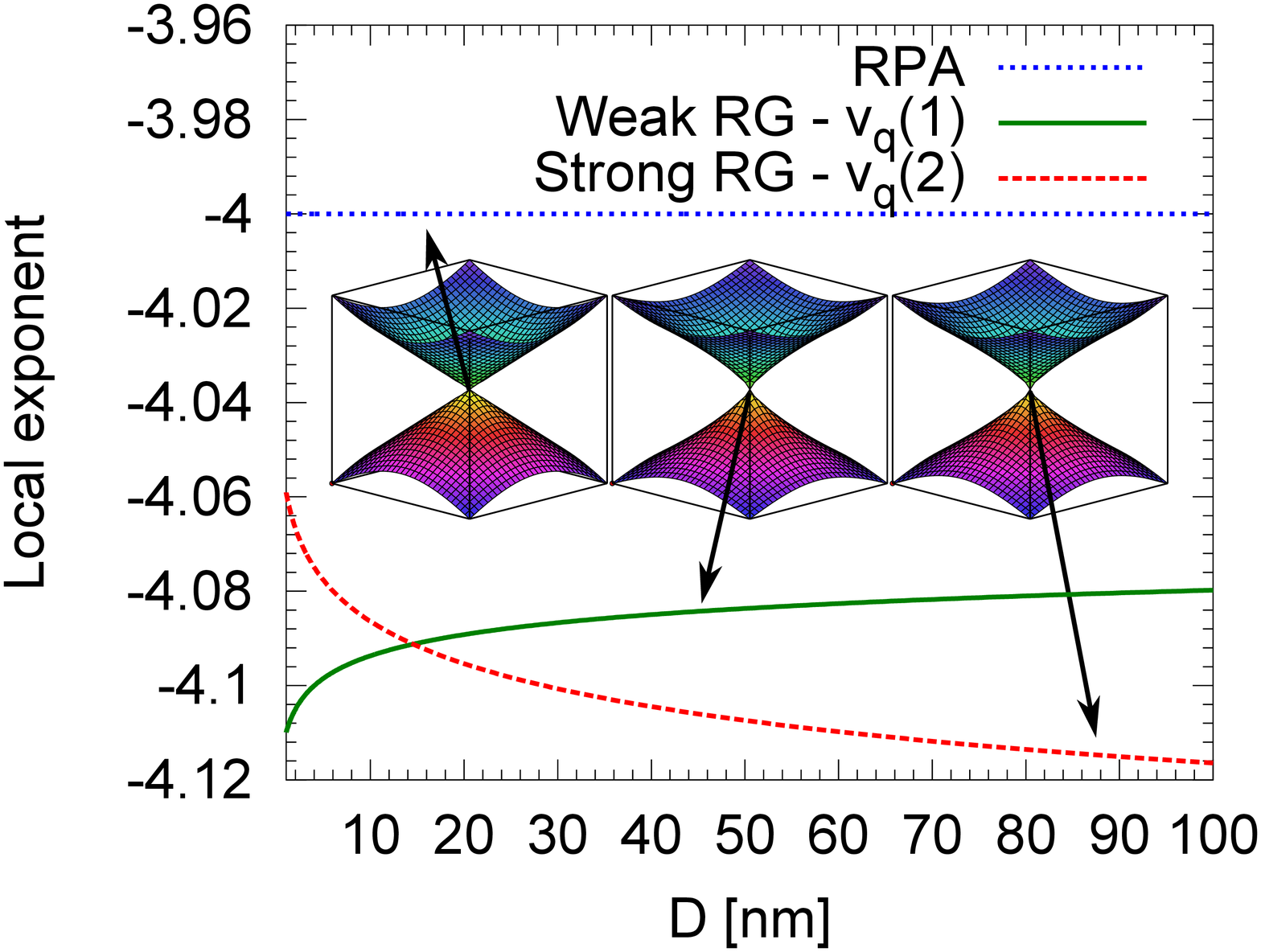}\\
\includegraphics[width=1.0\linewidth]{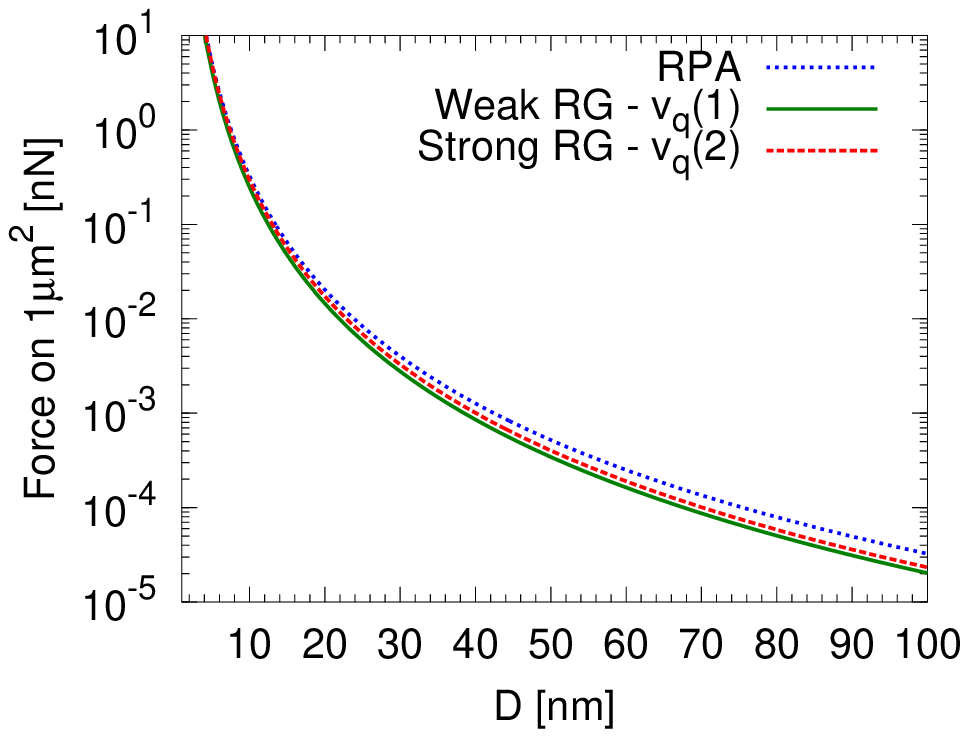}%
\caption{Plot of the local exponent $\frac{d\log{|F|}}{d\log{D}}$
  of the force (top, with bandstructure inset) and the force
  $F=\frac{d E^{(2)}}{d D}$ (bottom) using RPA [dotted line],
  $v_q^{(1)}$ [Eq.~\eqref{V1}, solid line] and
  $v_q^{(2)}$ [Eq.~\eqref{V2} with $\beta=2/\pi^2$, dashed line].
  Here we fix $v=10^5$m/s and $\Lambda=1/(1.42{\text{\AA}})$.}
\label{Fig3}
\end{figure}
Using this in Eq.~\eqref{ECPIntermediateD}, together with
Eq.~\eqref{V1} or \eqref{V2}
for the velocity, we find that the distance dependence
in the intermediate regime is basically $D^{-3}$ with only a modest
further dependence on distance via
$F\left(\frac{\pi}{2}\alpha_{q=1/D}\right)$ or
$\Ft\left(\frac{\pi}{2}\alpha_{q=1/D}\right)$.\footnote{Some care must
  be exerted when evaluating Eq.~\eqref{ECPIntermediateD} as the
  asymptotic renormalised perturbation approach for the corrections is
  not fully appropriate for larger $q$. In particular expression
  \eqref{V1} is invalid for large $q$ as $v_q$ can become negative.
  In our calculations we use instead
  $\alpha'_q=e^2/(\hbar\max[v,v_q])$ to avoid unphysical results.}

In Fig.~\ref{Fig3} we plot the force and its local exponent
[in $F\propto D^{-p(D)}$]
for the RPA energy $E^{(2)}_{\RPA}/A=-e^2/(32\pi D^3) F(\alpha)$ and the
vdW interaction calculated through \eqref{ECPIntermediateD}
using \eqref{V1} and \eqref{V2}.
Using the stretched graphite vdW energy formula of
Refs.~\onlinecite{Gould2009,Gould2013-Cones} yields remarkably similar
results for the equivalent in bulk graphite (up to a constant).
Naturally, the situation differs between the weak-coupling~\eqref{V1}
and strong-coupling~\eqref{V2} models of the renormalized velocity.
The interaction energy and force is qualitatively different from RPA
at large separations, and shows moderate quantitative differences
at intermediate separations, less than 25\% for $D<100$\AA.
Observation of such deviations from
the RPA will provide additional evidence of the many-body
renormalization of the fermion velocity.
We note that for intermediate values of $D$
($D\lesssim 20$nm), further modifications are
required even at the RPA level to account for departures
of the graphene electronic
bandstructure from a perfect infinite cone\cite{Gould2013-Cones},
or anisotropy\cite{Sharma2014}.

There have been some proposals (\cite{Kotov2012} and references
therein) that strong coupling could bring excitonic effects leading to
a gap. In that case the vdW energy might show the insulating $D^{-4}$
behavior as in Eq~\eqref{DMinus4Law}.
There seems to be little experimental evidence so far that
graphene can be an excitonic insulator, however.

\section{Experimental options}
\label{sec:Expt}

While our paper in the main concerns the theory of the
renormalization of the vdW interaction, it serves as one of several
motivating factors for a renewed experimental effort for direct
measurement of the vdW forces on high-quality graphene flakes.  These
measurements would also be needed to resolve existing controversies
about graphitic cohesion in general. In this section we briefly
discuss the current prospects for such measurements, and we hope
thereby to stimulate experimentalists in this direction.

While there have been a number of experiments%
\cite{Girifalco1956,Benedict1998,Zacharia2004,Liu2012}
that have
indirectly determined the binding energy of graphene planes, these
have produced a wide range of results, and most have relied on
questionable theoretical assumptions for their analysis, so that the
whole field is somewhat controversial\cite{Gould2013-Expt}. Purely
theoretical estimates have also varied widely
\cite{Dion2004,Hasegawa2004,Chakarova2006,Ziambaras2007,Hasegawa2007,%
Spanu2009,Lebegue2010,Drosdoff2010,Thrower2013,Bucko2013},
though recent very
large and relatively high-level (RPA\cite{Lebegue2010},
DMC\cite{Spanu2009}) calculations
are starting to show consistency. These same high-level types of
theory also predict the forces between nanostructures such as graphene
planes as a function of distance, showing very different results from
popular pairwise-additive-force theories, in the asymptotic region of
large separations [see Eqs~\eqref{DMinus3Law}, \eqref{DMinus3NewLaw}].

The force between graphenes at such asymptotic distances does not
appear to have been measured at all so far.  In view of this, as well
as the aforementioned binding-energy controversies, direct measurement
of such forces at all distances would be very desirable. As the
distant forces are relatively small, Atomic Force Microscopy
or a related NEMS oscillator approach would be the
preferred route.   For the basic RPA analysis of two cold undoped
graphene sheets of lateral dimension 1 micron, separated by 10 nm, the
predicted force is a few nanoNewton, well within AFM capabilities,
with larger forces for larger flake areas.

In the ideal case analyzed in the main text there are two
undoped, freestanding graphene sheets
at a temperature below 10K. Single sheets of high-quality graphene have
certainly been subjected to various measurements\cite{Knox2011}
and vdW forces
due to a single supported graphene sheet has been
seen\cite{Banishev2013}, but force experiments with two
freestanding sheets are rare or absent, and
will require some effort to ensure a reasonable degree of parallelism.

Perhaps a better short-term prospect is therefore
to measure the force between a single
freestanding sheet and its own ``vdW image'' in the surface of a bulk
metallic substrate. A very recent related experiment, using a metal
grating instead of graphene, and an $O(100{\mu}\text{m})$-radius gold
sphere
as the substrate to avoid parallellism issues, have been highly
successful\cite{Intravaia2013}. This experiment also demonstrated the
use of
co-located fiber optics for precise control of the separation $D$ down
to $< 200$nm. The theory of this geometry will be analyzed in detail
elsewhere, but is expected [see equation \eqref{eqn:GrMetal}]
to yield similar predictions to the ones in the main text.

The unavoidable corrugation of the freestanding graphene sheets should
not be a problem, as it is known experimentally not to affect
electronic properties substantially, and would only contribute a very
few percent uncertainty in the separation $D$ between the sheets, at the
separations of a few tens to hundreds of nanometers that are relevant
to the main text. This uncertainty would not affect the analysis of
the force power law proposed in the text, as one would aim for
measurements at $D$ values spanning an order of magnitude.

A more subtle problem is the likely existence of metallic n- and
p-type ``puddles'' on the undoped sheets\cite{Cheng2011}. Provided
that the sheets
are of sufficient quality that the puddles are disconnected objects of
typical spatial extent $\lambda$, we predict that they contribute an
``insulator-like'' vdW interaction energy varying with distance $D$ as
(const)$D^{-4}$ when $D>\lambda$, clearly distinguishable from the
lower powers
predicted in the text, arising from the un-doped non-puddle areas.
Recent experiments\cite{Cheng2011} suggest that $\lambda =
O(1\text{nm})$, so
that force experiments at $D = O(10\mbox{--}200\text{nm})$ would still
exhibit undoped-graphene properties.

Another experimental route would be to avoid separating the sheets in
their perpendicular direction, but rather to slide them off one
another in a surface-parallel direction. This may be easier
experimentally because of the recent centrifugation-based preparation
of high-quality micron-sized stacks of $2\mbox{--}5$ graphene
monolayers\cite{Chen2012}. These are spatially staggered
like a slipped deck of cards,
potentially allowing attachment of an AFM tip to the projecting edges
of individual sheets. The force during the entire lateral sliding
process, out to wide separation into disjunct coplanar sheets, would
be measured. The non-contact part of this force can be expected to
show effects from the coupling of long-wavelength electronic charge
fluctuations and hence renormalization effects in the graphene
polarizability, just as for the case of parallel sheets separated by
distance $D$ measured perpendicular to the sheets.  From the theory
point view, the analysis of this geometry is more difficult and has
not yet been attempted in detail.

The above considerations suggest that it will be possible to achieve a
much improved understanding of graphenic cohesive forces by direct
measurement.

\section{Conclusion}
In conclusion, we have shown that the low-temperature dispersion (vdW)
interaction between two infinite parallel non-doped graphene sheets is
significantly modified by many-body effects beyond the RPA.  Not only
is the interaction quantitatively reduced, but also its qualitative
asymptotic behavior is modified.  The main source
of the effect is the many-body
renormalization of the velocity of the massless Dirac Fermions.  This
renormalization has been the subject of many recent
investigations\cite{Gonzalez1994,Gonzalez1999,Li2008,Elias2011,Siegel2011}.
It is experimentally observed as a deformation of the Dirac cones near
the point of contact.  Our findings demonstrate that the same
renormalization manifests itself in the long-distance behavior of the
dispersion forces.

Direct measurement of the asymptotic graphene-graphene vdW interaction
could therefore distinguish between much-debated theory models of
electrons in graphene -- weak renormalization [Eq.~\eqref{AsyvdWRedByRenorm}],
strong renormalization [Eq.~\eqref{AsyvdWRedByRenorm2}]
and excitonic insulator [Eq.~\eqref{DMinus4Law}].
Such experiments in the asymptotic vdW region will be demanding but we
estimate (e.g.) a measurable  force of order nN
(see inset of Figure~\ref{Fig3}) between micron-sized
graphene sheets separated by $O(10\text{nm})$. Observation of the
vdW image force in a metal substrate could avoid the need for two
graphene sheets, and there are other possibilities
too (see Ref.~\cite{Chen2012} and second last paragraph of
Section~\ref{sec:Expt} above).
Very recent experiments\cite{Intravaia2013} support the general
feasibility of our proposals. We estimate that complications
due to graphene wrinkling and puddling\cite{Cheng2011} will not
destroy our effect. Indeed the
time is ripe for direct force measurements to clarify this and other
recent controversies\cite{Gould2013-Expt} over graphenic cohesion in
general.

\acknowledgments TG and JFD were supported by Australian Research
Council Grant DP1096240. GV was supported by NSF Grant
DMR-1104788. JFD appreciates hospitality at the University of
Missouri, the Donostia International Physics Centre, and the European
Theoretical Spectroscopy Facility.

\appendix
\section{van der Waals force between graphene and a metal bulk}
\label{app:Bulk}

Here we investigate the van der Waals force between a single
graphene layer and a metal bulk, which may be more appropriate
for likely successful experimental arrangements
(see Sec.~\ref{sec:Expt} below).
We predict that it will obey the same overall
power law as the response between two graphene layers,
with at most a logarithmic correction (as a function of $D$).
Similarly the dependence on the many-body beyond-RPA effects
is expected to be maintained. Here we offer evidence that
this is indeed the case.

We make use of the relationship\cite{Dobson2009-JCTN}
\begin{align}
\frac{E_{\vdW}}{A}=&\frac{\hbar}{8\pi^3}\int\d^2\vq\int_0^{\infty}\d u
\log(1-\zeta)
\label{eqn:EvdWB}
\\
\approx& -\frac{\hbar}{8\pi^3}\int\d^2\vq\int_0^{\infty}\d u \zeta
\end{align}
for the van der Waals potential between semi-infinite bulks and layers
interacting across a single surface. In the case of a layer of
graphene interacting with a bulk metal
\begin{align}
\zeta=&e^{-2qD}\frac{2\pi e^2}{q}\chi_{\text{Gr}}(q,iu)
\nonumber\\&\times
\frac{2\pi e^2}{q}
\int\d z\d z' e^{-q|z+z'|}\chi_{\text{Metal}}(q,z,z',iu)
\\
\equiv& C_{\text{Gr}}(q,iu) C_{\text{Metal}}(iu)e^{-2qD}
\end{align}
where $\chi_{\text{Gr}}(q,iu)$ is the interacting response of the graphene
layer and $\chi_{\text{Metal}}$ is the bulk response of the metal.
Here $C_{\text{Metal}}=\frac{\epsilon(iu)-1}{\epsilon(iu)+1}$ where
$\epsilon\approx 1+ \frac{\omega_p^2}{u^2}$ is the dielectric
function of the metal.

From equations \eqref{ChiFull} and \eqref{InteractingChi} we see that
$C_{\text{Gr}}\equiv \frac{-2\pi e^2}{q}\chi_{\text{Gr}}(q,iu)$
can be written as $C_{\text{Gr}}(x,\alpha_q)
\approx \frac{\pi \alpha_q/2}{\sqrt{1+x^2}+\pi \alpha_q/2}$
where
$x=u/(v_qq)$ and $\alpha_q=e^2/(\hbar v_q)$ varies slowly with $q$. Thus
\begin{align}
\frac{E_{\vdW}}{A}=&\frac{e^2}{4\pi^2 D^3}
\int \qbar^2 e^{-\qbar} \d \qbar \int_0^{\infty}\d x
\nonumber\\&\times
\frac{C_{\text{Gr}}(x,\alpha_{\qbar/(2D)})}{\alpha_{\qbar(2D)}}
\frac{1}{1 + \frac{v^2_{\qbar/(2D)}\qbar^2}{2D^2\omega_p^2}x^2}
\label{eqn:EvdWCP}
\end{align}
where we also used $\qbar=2qD$.
Clearly the form of $\alpha_q$ will have a substantial effect
on the asymptotic behaviour of the vdW potential, similar to
the bigraphene case.
Using $C_{\text{Gr}}<\frac{\pi/2\alpha}{x+\pi/2\alpha}$ and setting
$\alpha_{\frac{\qbar}{2D}}\approx \alpha_{1/D}$ gives
\begin{align}
\frac{E_{\vdW}}{A}<&\frac{e^2}{4\pi^2 D^3}
\int \qbar^2 e^{-\qbar} \d \qbar \int_0^{\infty}\d x
\nonumber\\&\times
\frac{\pi/2}{x+\pi/2\alpha_{1/D}}
\frac{1}{1 + \frac{v^2_{1/D}\qbar^2}{2D^2\omega_p^2}x^2}
\\
\approx& \frac{\alpha_{1/D}\log(D/D_0)}{D^3}
\label{eqn:GrMetal}
\end{align}
where $D_0\propto \frac{e^2}{\hbar\omega_p}=O(1\text{\AA})$
i.e. the asymptotic interaction has an extra logarithmic term compared
to the bigraphene case $k\alpha_{1/D}/D^3$. The prefactor
$\alpha_{1/D}$ ensures that the difference between different
renormalization scenarios is maintained.

One can, of course, perform the integral \eqref{eqn:EvdWCP} [or an
equivalent expression for \eqref{eqn:EvdWB}] numerically for a
given graphene velocity $v_q$ and dielectric frequency $\omega_p$.
However, the important asymptotic physics are clearer in the
bigraphene case tested in the paper, and the deviation caused
by the bulk metal is expected to be small at the
$O(10\text{--}100\text{nm})$ distances we propose investigating.


%

\end{document}